\documentclass[aps,twocolumn,showpacs,superscriptaddress,prl,10pt]{revtex4-1}
\usepackage{graphicx}
\usepackage{dcolumn}
\usepackage{bm}
\usepackage{color}
\usepackage{amsmath}
\usepackage{amsfonts}
\usepackage{epstopdf}
\usepackage{subfigure}

\begin{document}

\author{G. Bertaina}
\affiliation{Institute of Theoretical Physics, Ecole Polytechnique F\'{e}d\'{e}rale de Lausanne EPFL, CH-1015 Lausanne, Switzerland }

\author{E. Fratini}
\affiliation{School of Science and Technology, Physics Division, University of Camerino and CNISM, I-62032 Camerino, Italy}

\author{S. Giorgini}
\affiliation{INO-CNR BEC Center and Dipartimento di Fisica, Universit\`a di Trento, 38123 Povo, Italy}

\author{P. Pieri}
\affiliation{School of Science and Technology, Physics Division, University of Camerino and CNISM, I-62032 Camerino, Italy}
\affiliation{INFN, Sezione di Perugia, Perugia, Italy}

\title{Quantum Monte Carlo Study of a Resonant Bose-Fermi Mixture} 

\begin{abstract} 
We study a resonant Bose-Fermi mixture at zero temperature by using the fixed-node diffusion Monte Carlo method. We explore the system from weak to strong boson-fermion interaction, for different concentrations of the bosons relative to the fermion component.  We focus on the case where the boson density $n_B$ is smaller than the fermion density  $n_F$, for which a first-order quantum phase transition is found from a state with condensed bosons immersed in a Fermi sea, 
to a Fermi-Fermi mixture of composite fermions and unpaired fermions.
We obtain the equation of state and the phase diagram, and we find that the region of phase separation shrinks to zero for vanishing $n_B$.  
\end{abstract}

\pacs{67.85.Pq, 03.75.Ss,  03.75.Hh}
\maketitle

Let us consider a system of bosons and spinless fermions with a tunable short-range boson-fermion (BF) attraction. For weak attraction,  at sufficiently low 
temperature the bosons condense, while the fermions fill a Fermi sphere, and the BF interaction can be treated with perturbative methods \cite{Albus2002,Viverit2002}. For sufficiently strong attraction, bosons and fermions pair into molecules. In particular, for a fermion density $n_F$ larger than the boson density $n_B$, one expects all the bosons to pair with fermions. 
The boson condensate is then absent in such a regime, and the system should be described as a weakly interacting Fermi-Fermi mixture, one component consisting of molecules, with density $n_M=n_B$, and the other component of unpaired fermions, with density $n_U=n_F - n_B$. 

How does the system evolve at zero temperature  between the two above physical regimes? 
Several scenarios could be imagined in principle: 
(i) a continuous quantum phase transition, with the condensate fraction vanishing smoothly at a certain critical value of the BF coupling; 
(ii) a first-order quantum phase transition, with phase separation between a condensed phase and a molecular phase without condensate; 
(iii) the collapse of the system in the intermediate coupling region, with no stable state connecting the two different regimes.

The above question has been the object of increasing attention recently, especially in the field of ultracold trapped gases, where the interaction can be tuned by using Feshbach resonances
\cite{Chin2010}. In particular, for ``broad'' resonances, a Bose-Fermi mixture can be accurately described 
by a minimal set of  parameters: the scattering lengths $a_{BB}$ and $a_{BF}$ describing, respectively, the boson-boson (BB) and boson-fermion interaction, the boson and fermion densities $n_B$ and $n_F$, and their masses $m_B$ and $m_F$ (the short-range fermion-fermion interaction being 
negligible, due to Pauli exclusion).

Initial experiments \cite{Modugno2002,OspelkausC2006a} with ultracold Bose-Fermi mixtures  supported the collapse scenario, with the instability occurring already for moderate BF coupling. However, only a limited region of the parameter space was explored (e.g., a boson number $N_B$ considerably greater than the fermion number $N_F$ and nonresonant values of the scattering lengths). The tunability of the Bose-Fermi interaction was first demonstrated in a $^{40}$K$-^{87}$Rb mixture \cite{OspelkausS2006} and then successfully exploited to form  Feshbach molecules \cite{OspelkausC2006b,Zirbel2008,Ni2008}. Recently,  an isotopic $^{40}$K$-^{41}$K  mixture, exhibiting a broad Feshbach resonance to tune the BF interaction, was successfully cooled down to quantum degeneracy \cite{Wu2011}. The creation of Feshbach molecules  has been finally achieved very recently also  with $^{23}$Na$-^6$Li \cite{Heo2012} 
and $^{23}$Na$-^{40}$K \cite{Wu2012} mixtures, in the latter case observing lifetimes of the order of $100$ms, sufficient for the setup of many-body effects. The two above opposite regimes of a Bose-Fermi mixture have thus been explored already to some extent in experiments.  The intermediate (unitary) region, instead,  has remained inaccessible so far,  essentially because
of the large losses due to three-body recombination onto deep energy levels, favored by the presence of three-body (Efimov) bound states. Some control of these losses should, however, be achieved by working with small concentrations of bosons, the dominant recombination process being proportional to $n_B^2 n_F$, and by considering isotopic mixtures, for which Efimov states are relevant only very close to resonance \cite{Helfrich_2010}.

On the theoretical side, first studies \cite{Viverit2000,Yi2001,Roth2002}  were based on mean-field or perturbative approaches and focused mainly on the mechanical stability issue.  
The continuous quantum phase transition scenario was first put forward in  \cite{Powell2005}, within a ``two-channel model''  for the BF coupling  (actually relevant for ``narrow'' Feshbach resonances \cite{Giorgini2008,Chin2010}) and further explored, for a broad resonance,  in \cite{Fratini2010,Fratini2012}.
The alternative scenario of a first-order transition, with a rather vast region of phase separation occurring between the ``molecular'' region and the condensed one, was instead proposed in \cite{Marchetti2008,Ludwig2011}. A recent variational calculation has indicated finally that a sufficiently strong BB repulsion should prevent the collapse scenario \cite{Yu2011}.

In this Letter, we apply the fixed-node diffusion Monte Carlo (FN-DMC) method to the study of a resonant Bose-Fermi mixture. This numerical technique  yields an upper bound for the ground-state energy of the gas, resulting from an ansatz for the nodal surface of the many-body wave function that is kept fixed during the calculation (see Ref.~\cite{Reynolds1982} for details). 
We consider a three-dimensional homogeneous Bose-Fermi gas described by the Hamiltonian
\begin{multline}
H=-\frac{\hbar^2}{2m_F}\sum_{i=1}^{N_F}\nabla^2_i  -\frac{\hbar^2}{2m_B}\sum_{i^\prime=1}^{N_B}\nabla^2_{i^\prime}\\
+\sum_{i,i^\prime}^{N_F,N_B}V_{BF}(r_{ii^\prime})+\sum_{i^\prime<j^\prime}^{N_B}V_{BB}(r_{i^\prime j^\prime}) \;,
\label{hamiltonian}
\end{multline}   
where $i,j,...$ and $i^\prime,j^\prime,...$ label, respectively, the fermions and the bosons. We consider equal masses $m_B=m_F=m$ and model the interspecies BF interaction by using an attractive square-well potential with depth $V^0_{BF}$ and range $R_{BF}$, while the intraspecies BB interaction is modeled by a repulsive soft-sphere potential with height $V^0_{BB}$ and range $R_{BB}$.  

In order to eliminate any dependence on the range of the BF interaction potential, we take $R_{BF}$ such that $n_FR_{BF}^3=10^{-7}$ or, equivalently, $k_FR_{BF}=0.0181$ in terms of the Fermi wave vector $k_F=(6\pi^2 n_F)^{1/3}$. In this regime the only dependence on the BF interaction potential is given by the scattering length $a_{BF}$.
As is well known, $a_{BF}$ diverges and changes sign  when a two-body bound state appears (with binding  energy  $\varepsilon_B=-\hbar^2/m a_{BF}^2$ for $a_{BF}>0$). Deeper two-body bound states are absent in our model.
In the many-body system, the BF coupling strength is conveniently described in terms of the dimensionless parameter $g=(k_Fa_{BF})^{-1}$. 

For the BB repulsion, we set $R_{BB}=1.086 a_{BB}$ and take $\zeta \equiv k_Fa_{BB}=1$. Such a constant BB repulsion guarantees the stability for all considered values of the boson concentration $x=n_B/n_F$ and of the BF couplings across the resonance, preventing high local bosonic densities, which would favor the formation of clusters. We notice that our value of $\zeta$ is twice the critical value for mechanical stability found for $g=0$ in \cite{Yu2011}. 
Because of computational time constraints, an analysis of the  dependence on $\zeta$ of the results is beyond the scope of this Letter. This question is, however, definitively relevant for experiments with ultracold atoms, since typical values of $a_{BB}$ and $k_F$  correspond to values of  $\zeta$ smaller by at least one order of magnitude than the value considered here. In addition, with $\zeta=1$, we expect the specific choice of the repulsive potential to play a role for  $x\gtrsim 0.2$, based on previous studies of bosonic systems \cite{Giorgini1999,Pilati2008}.

Simulations are carried out in a cubic box of volume $L^3=N_F/n_F$ with periodic boundary conditions. We use a trial wave function  of the general form $\psi_T({\bf R})=~\Phi_S({\bf R})\Phi_A({\bf R})$. $\Phi_S$ is a positive function of the particle coordinates ${\bf R}=({\bf r}_1,\dots,{\bf r}_{N_F},{\bf r}_{1^\prime},\dots,{\bf r}_{N_B})$ and is symmetric under exchange of identical particles, while $\Phi_A$ satisfies the fermionic antisymmetry condition and determines the nodal surface of $\psi_T$.  The symmetric part is chosen of the Jastrow form $\Phi_S({\bf R})=\prod_{i,i^\prime}f_{BF}(r_{i i^\prime})\prod_{i^\prime j^\prime}f_{BB}(r_{i^\prime j^\prime})$, where the unprimed (primed) coordinates refer to fermions (bosons) and two-body correlation functions of the interparticle distance are introduced. In order to describe the Bose-Fermi and the Fermi-Fermi mixture, we use two different choices for $\Phi_A$.
The first choice (JS) is a usual Slater determinant for the bare fermions $\Phi_A^S({\bf R})={\cal A} \left( \psi_{k_1}(1)\psi_{k_2}(2)...\psi_{k_{N_F}}(N_F)\right)$, where ${\cal A}$ indicates the antisymmetrizer operator  and $\psi_{k_\alpha}(i)$ indicates a plane-wave state, with $k_\alpha=2\pi(n_{\alpha x}\hat{x}+n_{\alpha y}\hat{y}+n_{\alpha z}\hat{z})/L$ and $|k_\alpha|\le k_F$. The second choice (JMS) is the antisymmetrized product of a Slater determinant for the molecules and a Slater determinant for the unpaired fermions
\begin{equation}
 \Phi_A^{MS}({\bf R})=\left|\begin{matrix}
                        \varphi_{K_1}(1,1^\prime) & \cdots & \varphi_{K_1}(N_F,1^\prime) \\
			\vdots                   &\ddots & \vdots                  \\
			\varphi_{K_{N_M}}(1,{N_M}) &\cdots & \varphi_{K_{N_M}}(N_F,{N_M}) \\
			\psi_{k_1}(1)          &\cdots & \psi_{k_1}(N_F)     \\
			\vdots                           &\ddots & \vdots                  \\
			\psi_{k_{N_U}}(1)        &\cdots & \psi_{k_{N_U}}(N_F)
                       \end{matrix}\right|
\label{eq:psiA-MS}
\end{equation}
where the molecular orbitals are defined as $\varphi_{K_\alpha}(i,i^\prime)=f_b(|{\bf r}_i-{\bf r}_{i^\prime}|)\exp{(i {\bf K}_\alpha ({\bf r}_i+{\bf r}_{i^\prime})/2)}$, which consist of the relative motion orbitals $f_b$ times the molecular center-of-mass plane waves with $|K_\alpha|\le K_M$, being $n_M=K_M^3/6\pi^2$, while for the unpaired fermions $|k_\alpha|\le k_U$, being $n_U=k_U^3/6\pi^2$. The functions $f_b$, as well as $f_{BB}$ and $f_{BF}$, are chosen to be the solutions of the appropriate two-body problems, modified at long distance to comply with periodic boundary conditions. We notice that the wave function \eqref{eq:psiA-MS} is not symmetric under exchange of bosons. It is the analog of the Nosanow-Jastrow wave function \cite{Nosanow1964,Hansen1968}, which has been successfully used in quantum Monte Carlo studies of the equation of state of solid $^4$He \cite{Whitlock1979,cazorla2009}.

\begin{figure}[ptb]
\begin{center}
\includegraphics[width=0.95\columnwidth]{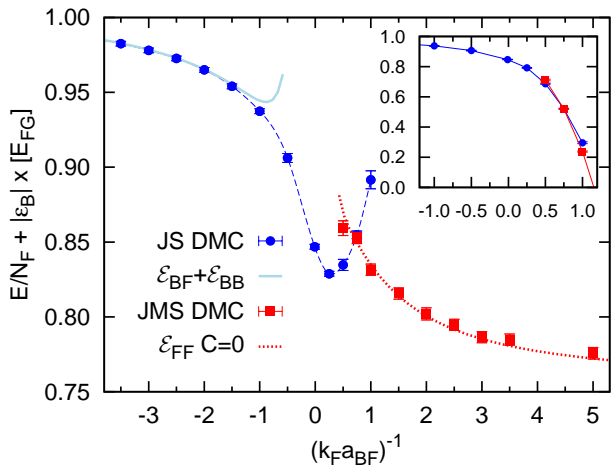}
\caption{\small{Energy of a BF mixture at $x=0.175$ and $\zeta=1$, with the contribution of the bare binding energy of the molecules subtracted for $a_{BF}>0$. Circles: JS FN-DMC results. The solid line corresponds to  eq.~\eqref{eq:ebfw} and the dashed line is a guide to the eyes. Squares: JMS FN-DMC results. The dotted line corresponds to eq.~\eqref{eq:effsm} with $M^*$ and $\alpha$ from \cite{Combescot2009a} and $C=0$. Inset: Energy without subtracting the bare binding energy.}}\label{fig:eos}
\end{center}
\end{figure}

In  Fig.~\ref{fig:eos}  we report the FN-DMC results for the total energy of the mixture at a small boson concentration $x=0.175$ as a function of the interaction parameter $g$ in units of the energy per particle of the free Fermi gas $E_{\text{FG}}=3\hbar^2k_F^2/10m=3\varepsilon_F/5$, where $\varepsilon_F$ is the Fermi energy. 
We perform calculations with $N_F=57$, $N_B=10$ for the JS nodal surface ($\Phi_A^S$) and with $N_F=40$, $N_B=7$ with the JMS wave function ($\Phi_A^{MS}$), in order to have almost equal bosonic concentrations. For the JS (JMS) nodal surface  finite-size effects are considerably reduced by using closed shells for the number of fermions (molecules and unpaired fermions)  and using Fermi liquid theory. The energy difference between the finite and infinite systems is assumed to be the same as for the noninteracting case
(see~\cite{Ceperley1987}). We use this correction also to assess the error bars, on top of the statistical error. 

In the weak-coupling limit we recover the perturbative results of Refs.\cite{Viverit2002,Albus2002}, 
which can be further expanded in powers of $x$, leading to the energy functional $E=N_FE_{\text{FG}} ({\cal E}_{BF} + {\cal E}_{BB})$, where
\begin{equation}
  {\cal E}_{BF} = 1 + \frac{20}{9\pi g} x \left( 1+\frac{1}{\pi g} \right) + \frac{10 \zeta x^2}{9 \pi g^2}\left(1+\frac{4}{\pi^2}\right)  \label{eq:ebfw}
\end{equation}
while ${\cal E}_{BB}=\frac{10 \zeta x^2}{9 \pi} \left(1+ \sqrt{x}\zeta^{3/2} \frac{128}{15\pi\sqrt{6\pi}}\right)$ corresponds to the energy of a weakly interacting Bose gas \cite{LeeHuangYang1957}.

\begin{figure}[ptb]
\begin{center}
\includegraphics[width=0.95\columnwidth]{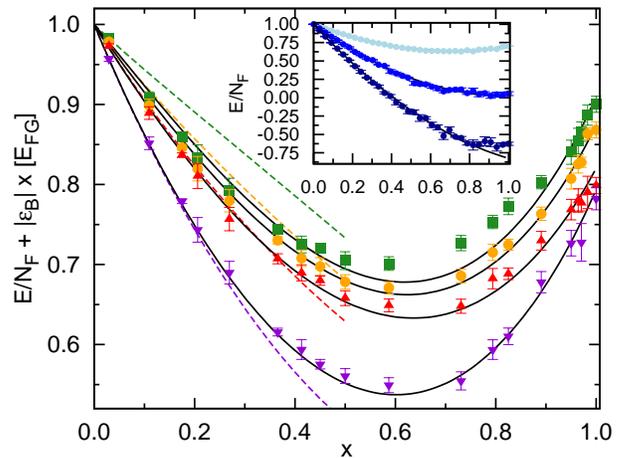}
\caption{\small Energy as a function of $x$, for $\zeta=1$. Main figure, from top to bottom: Energy of the FF mixture at $g=0.6$, $g=0.75$, $g=1$, and $g=5$, with the bare binding energy of the molecules subtracted. Dashed lines: Equation~\eqref{eq:effsm} with $C=0$, $M^*=2m$ and $\alpha=1.18$. Solid lines: Best fit using eq.~\eqref{eq:effsm} with $M^*$ and $\alpha$ from \cite{Combescot2009a}. Inset, from top to bottom: Energy of the polaronic phase at $g=0$,  $g=0.5$, and $g=0.75$. Solid lines: Best fit using Eq.~\eqref{eq:ebfp}.}\label{fig:x}
\end{center}
\end{figure}

More generally, the condensed phase can be described in terms of  a {\it polaronic} picture, where bosons are dressed by fermions. These polarons are characterized by an effective  binding energy $A$ and an interaction term $F$. Similarly to \cite{Yu2011}, one can thus introduce the following polaronic equation of state (EOS) holding in the limit $x\ll1$:
\begin{equation}\label{eq:ebfp}
 E_{\text{pol}}=N_FE_{\text{FG}}\left[1- A(g) x + F(g,\zeta)x^2\right]\;,
\end{equation}
where $A(g)=-(\mu_B/E_{\text{FG}})_{x\to 0}$ is calculated within a T-matrix framework \cite{Combescot2007,Fratini2010},  $\mu_B$ being the chemical potential of the bosons, while $F(g,\zeta)=\frac{10 \zeta }{9 \pi}\left(1 + D(g,\zeta)\right)$. An analogous $(x^2)$ interaction term has been considered in the context of polarized Fermi gases \cite{Mora2010, Yu2010, Giraud2012}.
In order to precisely evaluate the interaction coefficient $D$ of the polaronic EOS, we vary the concentration of the bosons in the relevant regime of couplings $0\le g\le1$, and we fit the coefficient  directly from the FN-DMC simulations. Results for the polaronic branch are shown in the inset in Fig. \ref{fig:x}; in Table~\ref{tab:cndn}, we report the fitted values for $D$. The agreement with the polaronic EOS is rather good even at large concentrations. Some discrepancies start to appear at large $x$ for $g=0.75$.

\begin{table}[pb]
\caption{Best fitted values of the $D(g,\zeta)$ coefficient of the polaronic EOS \eqref{eq:ebfp} and of the $C(g,\zeta)$ coefficient of the molecular EOS \eqref{eq:effsm} for $\zeta=1$.}
\begin{tabular}{c|c||c|c}
$g$ & $D$ & $g$ & $C$\\ 
\colrule
0.00 &  0.99(1)  & 0.60 & 12.39(10)   \\
0.25 & 1.33(5)   & 0.75 & 3.37(2)   \\
0.50 &  1.75(5)  & 1.00 & 1.54(2)   \\
0.75 &  1.95(5) &   5.00 & 0.60(1)    \\
1.00 &  1.25(1)  &   &   
\label{tab:cndn}
\end{tabular}
\end{table}

In Fig.~\ref{fig:eos}, we compare the FN-DMC results with the JMS wave function to the energy functional $E_{\text{mol}}=N_F E_{\text{FG}}(-10 g^2 x/3 +{\cal E}_{FF})$, where the first contribution comes from the bare binding energy of the molecules and
\begin{multline}\label{eq:effsm}
  {\cal E}_{FF} =  \frac{m}{M^*(g)}[1+xC(g,\zeta)]x^{5/3} + (1-x)^{5/3} \\
 + x(1-x)\frac{5 \alpha(g)}{3\pi g}\;,
\end{multline}
which is expected to hold for large values of $g$. Here, the first term corresponds to the kinetic energy of the molecules whose effective mass is given by $M^*(g)$, taken from the analytic treatment of  \cite{Combescot2009a} for a single molecule in a Fermi sea, corrected by a term proportional to the coefficient $C$ for finite values of $x$. This higher order $x^{8/3}$ contribution could also embody a $p$-wave interaction between the molecules, which is expected to be significant for $\zeta=1$. The second and third terms correspond instead to the kinetic energy of the unpaired fermions and to the interaction energy of the two fermionic components which, at the level of mean-field theory, is proportional to the ratio $\alpha=a_{ad}/a_{BF}$ of the atom-dimer to the BF scattering lengths; the $g-$dependence of this coefficient is taken from \cite{Combescot2009a} and in the strong-attractive (molecular) limit correctly reduces to the value $\alpha=1.18$ obtained from the solution of the three-body 
problem~\cite{Skorniakov1956,Petrov2004}. At the small value of $x=0.175$ shown in Fig.~\ref{fig:eos}, the FN-DMC results  compare well with the EOS~(\ref{eq:effsm}) with $C=0$.

Analogously to the polaronic branch, we  perform simulations using the JMS wave function for $g\ge0.6$ and different concentrations of the bosons. Results of the molecular FF mixture are shown in Fig. \ref{fig:x}. For the three largest values of $g$ we find that the EOS in Eq.~(\ref{eq:effsm}), including the correction to $M^*$ linear in $x$, works well up to $x=1$. The corresponding best fitted values of the coefficient $C$ are reported in Table~\ref{tab:cndn}. For $g=0.6$, our results start showing some deviations from the functional form (\ref{eq:effsm}) in the regime of intermediate concentrations $0.5\le x\le0.9$. For even smaller values of $g$ a number of effects worsen the agreement with the FN-DMC data: (i) The molecular effective mass from \cite{Combescot2009a} diverges for $g\simeq 0.5$, indicating that a molecular picture is not valid anymore, (ii) beyond mean-field interaction terms for the FF mixture are probably relevant, and (iii)  the composite nature of the molecules should start to play a major role.

\begin{figure}[ptb]
\begin{center}
\includegraphics[width=0.95\columnwidth]{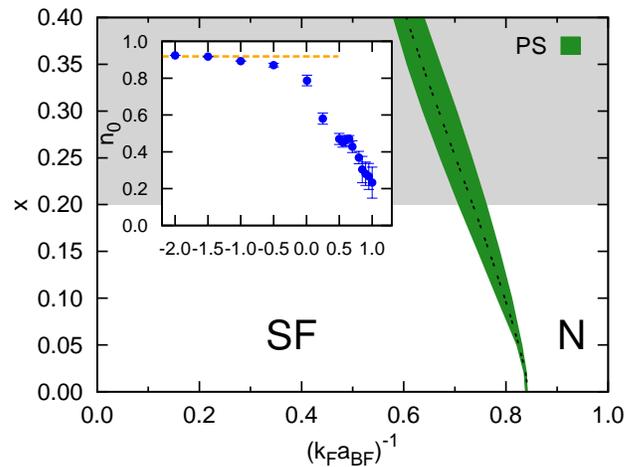}
\caption{Phase diagram in the $x$-$g$ plane, at $\zeta=1$. The dotted curve  indicates the energy crossing between the polaronic (SF) and FF molecular (N) phases. It lies  in the middle of a  phase separation (PS) region. The gray area above $x\simeq0.2$ indicates the nonuniversal region (for  $\zeta=1$). Inset: Bosonic condensate fraction $n_0$ for $x=0.175$ in the polaronic phase. Circles: Extrapolated results from FN-DMC and VMC simulations. Dashed line: Weakly-interacting Bose gas result $n_0^B=1-(8/3)\sqrt{x \zeta^3/6 \pi^3}$. For $g>0.75$ the unbiased estimate of $n_0$ is no longer reliable, resulting in large error bars.}
\label{fig:phase}
\end{center}
\end{figure}

We pass now to discuss the bosonic condensate fraction $n_0=N_0/N_B$. For the polaronic phase one can determine $n_0$ by calculating the unbiased long-range tail of the bosonic one-body density matrix from FN-DMC and variational Monte Carlo (VMC) simulations (see, {\it e.g.}, Ref.~\cite{Kolorenc2011}). The results for $x=0.175$ are reported in the inset in  Fig.~\ref{fig:phase} and show a constant decrease of $n_0$ from the weakly interacting regime, where the only contribution to depletion comes from the BB repulsion, to the strongly interacting regime, where the BF interaction dominates. In this region, however, the FN-DMC and VMC results for $n_0$ start to differ significantly, preventing an accurate unbiased determination of this quantity. As a consequence, we cannot assess whether $n_0$ decreases to zero by following the polaronic branch deeper on the molecular side or, as it is more reasonable, remains always finite. In the molecular phase, which for large values of $g$ is well described by the FF mixture of Eq.~\eqref{eq:effsm}, one expects $n_0=0$.

Figure \ref{fig:phase} finally presents  the phase diagram in the $x$-$g$ plane, for small concentration $x$, where we are confident in the validity of the energy functionals (\ref{eq:ebfp}), (\ref{eq:effsm}) describing, respectively, the superfluid (SF) polaronic and the normal (N) FF molecular phase. The dotted curve corresponds to the energy crossing between the two phases.  The two homogeneous phases are separated by  a phase separation region,  obtained by finding the global minimum of the energy functional  $v_PE_{\text{pol}}(g_P, x_P )+ (1-v_P) E_{\text{mol}}(g_M,x_M)$ with respect to the fractional volume $v_P$ of the polaronic phase, and the local couplings and concentrations $g_P, x_P, g_M, x_M$, at  fixed global particle numbers.  
 (We have checked that, at equilibrium, the resulting local couplings and concentrations lie in the respective regions of validity of the two energy functionals.) 

Our FN-DMC calculations support  the scenario of a first-order quantum phase transition, with a  narrow phase separation region intervening between the condensed polaronic  and FF molecular phase. The phase separation region shrinks to zero in the limit $x \to 0$, where the transition line tends to the critical coupling for the polaron-molecule transition, previously studied in the context of polarized Fermi gases \cite{Prokofev2008,Combescot2009a,Punk2009}. However, contrary to Fermi gases, the polaron-molecule transition is not masked by a large phase separation region with finite width even in the limit $x\to0$ \cite{PIlati2008a}. The physical reason for this is the different contribution of molecules to the EOS: In polarized Fermi gases, molecules are composite bosons and feel a repulsion mediated by the dimer-dimer scattering length; here they are composite fermions and feel the Pauli repulsion. The polaron-molecule transition is thus connected continuously to a quantum phase transition occurring at  {\em finite} boson concentration, a result which does not depend on the value of $\zeta$, provided the polaronic phase is mechanically stable, as can be seen from a small $x$ expansion of the energy functionals. 
We note finally that such a conclusion would still be valid even if the first-order quantum phase transition found here was replaced by a continuous phase transition, a scenario which we cannot completely exclude as our results depend on the choice of the competing states. Nevertheless, the good agreement of our results with controlled expressions in different limits and with established results for the polaron-molecule transition, together with the quite narrow phase separation region between the two phases considered in the present study, indicate that any improved interpolating trial nodal surface could possibly be relevant only in a narrow region of the phase diagram.

\begin{acknowledgments}
 We thank V. Savona for many helpful discussions and constant support throughout this project. P.P. acknowledges financial support from EPFL academic visitor program.
\end{acknowledgments}


%

\end{document}